\documentclass[preprint,showpacs]{revtex4}
\usepackage{amsmath,amsfonts,amssymb,dcolumn,bm}
\usepackage[english]{babel}
\usepackage[dvips]{graphicx}
\begin{document}

\title{Shock creation and particle acceleration\\ driven by plasma expansion into a rarefied medium}
\author{G. Sarri}
\affiliation{Centre for Plasma Physics, The Queen's University of Belfast, Belfast, BT7 1NN, UK}
\author{M. E. Dieckmann}
\affiliation{VITA ITN, Linkoping University, 60174 Norrkoping, Sweden\\}
\author{I. Kourakis}
\affiliation{Centre for Plasma Physics, The Queen's University of Belfast, Belfast, BT7 1NN, UK}
\author{M. Borghesi}
\affiliation{Centre for Plasma Physics, The Queen's University of Belfast, Belfast, BT7 1NN, UK}

\begin{abstract}
The expansion of a dense plasma through a more rarefied ionised medium is a phenomenon of interest in various physics environments ranging from astrophysics to high energy density laser-matter laboratory experiments.
Here this situation is modeled via a 1D Particle-In-Cell simulation; a jump in the plasma density of a factor of 100 is introduced in the middle of an otherwise equally dense electron-proton plasma with an uniform proton and electron temperature of 10eV and 1keV respectively.
The diffusion of the dense plasma, through the rarified one, triggers the onset of different nonlinear phenomena such as a strong ion-acoustic shock wave and a rarefaction wave. Secondary structures are detected, some of which are driven by a drift instability of the rarefaction wave.
Efficient proton acceleration occurs ahead of the shock, bringing the maximum proton velocity up to 60 times the initial ion thermal speed.
\end{abstract}

\maketitle

\section{Introduction}
\verb|  |The impact of a high energy-density laser pulse on a solid target results in the evaporation of the target material and the sudden ionisation, driven by the X-rays generated during the main interaction, of the surrounding low-density gas.  The ablated plasma expands under its own pressure through the dilute plasma triggering the creation of a bunch of non-linear structures such as collisionless shocks \cite{Forslund}. The possibility of generating collisionless shocks in laboratory is of extreme importance because it permits us to study their dynamics in a controllable manner. A  better  understanding  of such  shocks  is  not  only  relevant  for the  laser- plasma experiment as such and for inertial confinement fusion experiments. It can also provide further insight into the dynamics of solar system shocks \cite{Lee} and the nonrelativistic astrophysical shocks, like the supernova remnant shocks \cite{Kulsrud} and the consequent bursts of electromagnetic waves and accelerated particles that they induce \cite{McClements,Uchiyama,Aharonian}.\\
\verb|  |Such structures, that tend to have a characteristic width of the order of the plasma Debye length, have been recently detected in laser-plasma experiments \cite{Romagnani,Sarri1} employing the well-established proton imaging technique \cite{Sarri2}. These structures have been detected during the interaction of a ns laser beam with an average intensity of $10^{14}\div10^{15}$W/cm$^2$ with metallic targets; these beams ablate the surface of the solid target and, as a result, a warm plasma, with an average electron density of $\approx10^{18} \div 10^{19}$cm$^{-3}$ expands under its own pressure together with the isotropic generation of a black-body like X-ray spectrum. Both at the front and rear surface of the target, the X-rays suddenly ionise the background gas creating a rarefied plasma which typical densities are of the order of $10^{13}\div10^{15}$cm$^{-3}$; plasma structures of both electronic and ionic nature have been detected to propagate through this medium.\\
\verb|  |This renewed set of experimental data provides then a significant motivation for related numerical simulations especially concerning the shock creation mechanism.\\
\verb|  |The expansion of a warm plasma, dragged by the charge imbalance left behind by the hot electron acceleration, has been numerically studied in a number of papers.
However, most of these works rely either on the approximation of a self-similar expansion \cite{Dorozhkina}, or to Maxwellian distributed electrons with a time-dependent temperature \cite{True,Mora1}.
Deviations from the Maxwellian distribution of electrons have been highlighted only recently \cite{Mora2,Grismayer} and have been shown to play a significant role in the plasma dynamics giving, as an example, a justification of the energy increase of the forward accelerated proton beam.\\
\verb|  |Here we present the results obtained via a 1-dimensional Particle-In-Cell (PIC) simulation of the expansion of a dense electron-proton plasma into a tenuous one. We introduce a sharp density jump of the order of 100 in the middle of an otherwise equally dense plasma with protons and electrons both following Maxwellian distributions with a zero mean speed and with temperatures of 10eV and 1keV respectively throughout the entire simulation box. This difference of temperature between the two species is consistent with a partial thermalisation of the plasma as it is expected in the early, transient expansion of the plasma, time at which such non-linear structures are expected to form.
It has to be noted that such a density jump is considerably less than the one effectively experienced in laser-plasma laboratory experiments; nevertheless, in this case the transition between the two densities will not be sharply defined but will decrease gently with a rate that would be comparable, if not less, to the one artificially introduced here in the simulation initial conditions.\\
\verb|  |The structure of the paper is the following. In Section 2 we describe the PIC method and its governing equations and we give the initial conditions for the simulation parameters.
In Section 3 we proceed to analyse the preliminary stage of expansion of the plasma. The different dynamics of the two species of the plasma immediately induces a net electric field at the junction point, first signature of the onset of a rarefaction wave. Later in time, this wave is disrupted by the creation of a strong collisionless shock that propagates at a speed of the order of the ion-acoustic speed. During their diffusion, the electrons are effectively seen to gradually switch from a Maxwellian distribution to a flat-top one, as predicted in \cite{Mora2,Grismayer}.
In Section 4 we will then discuss the results from the late time evolution; while the electrons preserve their smooth diffusion, secondary waves are created in proximity of the rarefaction wave, some of which are triggered by the onset of a drift instability \cite{Hoh}. Moreover, the protons ahead of the shock are accelerated up to $\approx36$keV (i.e. 3600 times the initial proton temperature) corresponding to $\approx0.15$ times the electron thermal speed.\\
\verb|  |Finally in Section 5 we will give a brief summary of the relevant features observed.\\
\section{Particle-In-Cell Approach and Initial Conditions}
\verb|  |A Particle-In-Cell (hereafter simply referred as PIC) code approximates a plasma as an ensemble of computational particles (CPs) each of which represents a phase space volume element. The dynamics of each CP is determined by the Lorentz force played by a spatially and temporally dependent distribution of electric \textbf{E} and magnetic \textbf{B} fields. Both fields are self-consistently evolving following the Maxwell equations with the macroscopic plasma current \textbf{J}, which is the sum over the partial currents of all CPs.\\
\verb|  |The initial conditions of our simulation specify a plasma with a sharp density jump, separating the space in two different regions to be henceforth referred to as ''plasma 1'' and ''plasma 2'': plasma 1 consists of electrons and protons with density $n_1$ and temperature 1keV and 10eV respectively, whereas plasma 2 consists of electrons and protons of density $n_2 = n_1 / 100$ and the same temperatures as above. All the species have an initial Maxwellian velocity distribution centered at zero.\\
\verb|  |We normalise the solved equations with the help of the number density $n_2$, the plasma frequency $\omega_{p2} = (n_2 e^2/ m_e \epsilon_0)^{1/2}$ and the Debye length $\lambda_D = v_e/\omega_{p2}$ (where $v_e$ represents the electron thermal speed).
This choice is justified by the fact that, for this class of phenomena, the most interesting features occur in the low density plasma \cite{Romagnani}.
The normalisation of the relevant physical quantities leads to define then: $E_p = \omega_{p2} v_e m_e E/e$, $B_p = \omega_{p2} m_e B/e$, $J_p = e v_e n_2 J$, $\rho_p = e n_2 \rho$, $x_p = \lambda_D x$, $t_p = t/\omega_{p2}$, where the subscript $p$ stands for quantities in physical units. In this normalisation the charge $q$ is 1 for the protons and -1 for the electrons while the mass is $m_e = 1$, $m_p = 1836$. Following this normalisation, the system of equations that the code solves is formed by the four Maxwell equations:
\begin{equation}
\nabla\times\textbf{B} = \tilde{v}_e^2(\partial_t \textbf{E} + \textbf{J})
\end{equation}
\begin{equation}
\nabla\times\textbf{E} = -\partial_t \textbf{B}
\end{equation}
\begin{equation}
\nabla\cdot\textbf{E} =  \rho
\end{equation}
\begin{equation}
\nabla\cdot\textbf{B} = 0,
\end{equation}
where $\tilde{v}_e = v_e/c$, plus the Lorentz force:
\begin{equation}
m_i\partial_t \textbf{v}_i =q(\textbf{E}(x_i) + \textbf{v}_i\times\textbf{B}(x_i))
\end{equation}
and the equation of motion:
\begin{equation}
\frac{d}{dt}x_i = v_{i,x}
\end{equation}
The Lorentz force is solved for each CP of index $i$, position $x_i$ and velocity $\textbf{v}_i$.\\
\verb|  |Each species of the dense cloud is represented by $1.2\cdot10^8$ CPs and each of the tenuous species by $1.2\cdot10^6$ CPs. The interpolation schemes used by the code to infer the electromagnetic fields from the grid to the particle position is described in \cite{Dawson}, while a detailed description of the method used by the code can be found in \cite{Eastwood}.\\
\verb|  |In this paper we restrict our attention to one spatial dimension ($x$) with periodic boundary conditions and a purely electrostatic regime ($B=0$). Given that the length of the simulation box is $L$, plasma 1 occupies the first half (i.e. $-L/2<x<0$) while plasma 2 occupies the second half ($0<x<L/2$). The box length $L=4000$  is divided into 30000 cells. Such a large simulation box is required in order to minimise depletion effects on the hot electrons caused by the plasma expansion. The periodic boundary conditions imply that a practically identical second plasma expansion takes place at $x=L/2$, which we do not consider here.\\
\verb|  |The plasma evolution in time will be monitored, throughout the paper, by looking at the electron and proton phase space plots together with the relative spatial distribution of the electric field modulus. The choice of plotting the modulus of the electric field, instead of its real value, is simply dictated by the necessity of improving the otherwise poor signal-to-noise ratio. Due to the small number of particle per cell that PIC simulations can ensure, only a low statistical representation of the plasma can be achieved; this is mainly translated in rapid fluctuations, with constant amplitude, of the electric fields. These fluctuations, which tend to spatially extend over few Debye lengths, can be smoothed out by taking the electric field modulus, without modifying the data interpretation.
\newpage
\section{Initial Development}
\begin{figure*}[h!]
\begin{center}
\includegraphics[width=0.9\columnwidth]{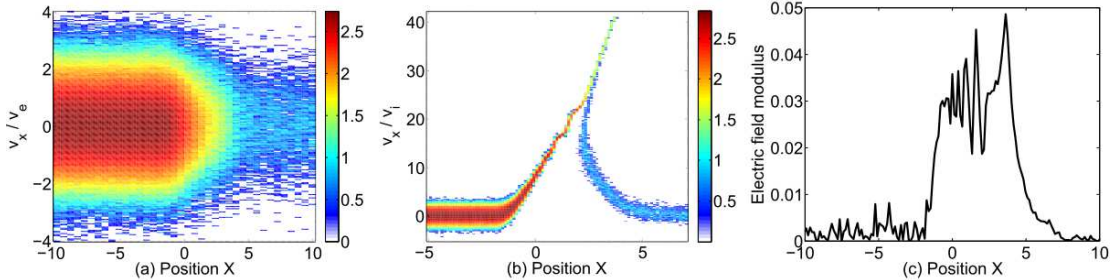}
\caption{(color online) The plasma state at the time $t_1 = 56$: Panel (a) displays the electron distribution and (b) the proton distribution. Panel (b) shows also the time evolution of the protons throughout the simulation (enhanced online). The 10-logarithmic color scale corresponds to the number of computational particles. The electrostatic field modulus is shown in panel (c).\label{fig2}}
\end{center}
\end{figure*}
As discussed above, the initial conditions of our simulation involve a density jump located at $x = 0$; the electrons of the high density part start then immediately to diffuse, guided by their own pressure into the low density part. Since the more massive protons can not keep up with this motion, a spatial charge imbalance occurs inducing a net electrostatic field with an approximately constant value in the region $-2\leq x \leq4$ (Fig. \ref{fig2}.c).
Such an electric field accelerates the protons close to the initial density jump that, already at $t_1 = 56$, reach velocities of $v_x \approx 40 v_i$, where the proton thermal speed $v_i = 0.1 / (1836)^{1/2}$ (see Fig. \ref{fig2}.b). The protons are accelerated further as the time animation of the proton distribution in Fig. \ref{fig2}.b evidences. Already at this very early time, the phase space distribution starts to show a bending of the protons of plasma 2, which is a first signature of the shock creation. The accelerating protons of plasma 1 are confined to a narrow beam with a mean speed that increases approximately linearly between $-2<x<4$, which is typical for a rarefaction wave. In the meanwhile, nothing relevant looks to occur in the electron phase space during the expansion towards the half plane $x>0$ (Fig. \ref{fig2}.a).\\
\begin{figure*}[h!]
\begin{center}
\includegraphics[width=0.9\columnwidth]{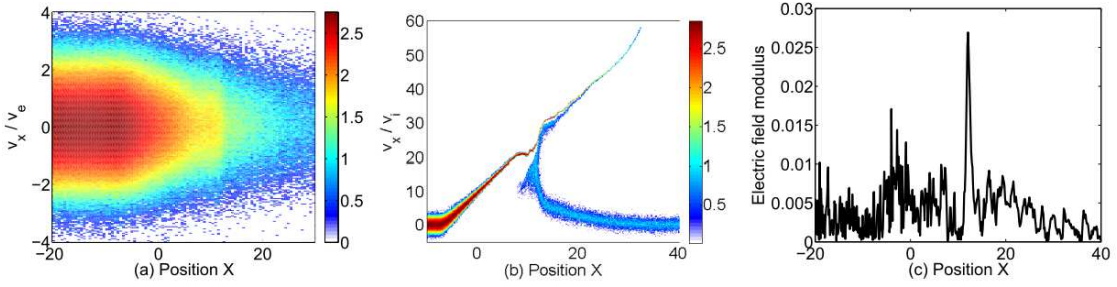}
\caption{(color online) The plasma state at the time $t_2 = 288$: Panel (a) displays the
electron distribution and (b) the proton distribution. The 10-logarithmic
color scale corresponds to the number of computational particles. The
electrostatic field modulus is shown in panel (c).\label{fig3}}
\end{center}
\end{figure*}
\verb|  |At $t_2 = 288$ the bending of the proton distribution of plasma 2 has become more pronounced; the proton phase space starts to show a diagonal structure, starting at $x \approx -5$ and extending up to $x \approx 32$. This structure is related to the onset of a rarefaction wave and it is disrupted by a sudden increase of the proton velocity located at $x\approx 10$ (Fig. \ref{fig3}.b), first signature of the formation of a collisionless ion-acoustic shock. A tiny downstream region (i.e. the flat plateau in the proton phase space at $9\leq x \leq12$) keeps the left part of the rarefaction wave separated from the forming shock. The rarefaction wave is associated with a spatially broad electric field distribution, that is responsible for both, the powerful acceleration that the protons experience in the same interval and the gradual decrease of the electron density. Its amplitude modulus $\le 0.01$ is less than that in Fig. \ref{fig2}.c and it has increased, between $t_1=56$ and $t_2=288$, the maximum proton velocity up to $v\approx60 v_i$. This acceleration is driven by the electrons: the expanding electrons will move to the right, leaving behind a charge imbalance that maintains the electric field that tries to hold them back. The charge imbalance is evident if we look at the density plots in Fig.\ref{densityplots}. At a very first stage of the expansion ($t_1=56$, Fig. \ref{densityplots}.a), the more massive protons cannot keep up with the electron expansion, thus setting a net electric field (i.e. the one plotted in Fig. \ref{fig2}.c). Later in time ($t_2=288$, Fig. \ref{densityplots}.b), the charge imbalance has significantly reduced, explaining the decrease of the amplitude of the broad electric field located at $-5 \leq x \leq 8$ (compare Figs. \ref{fig2}.c and \ref{fig3}.c). This is the region associated with the downstream part of the rarefaction wave.
\begin{figure*}[h!]
\begin{center}
\includegraphics[width=0.8\columnwidth]{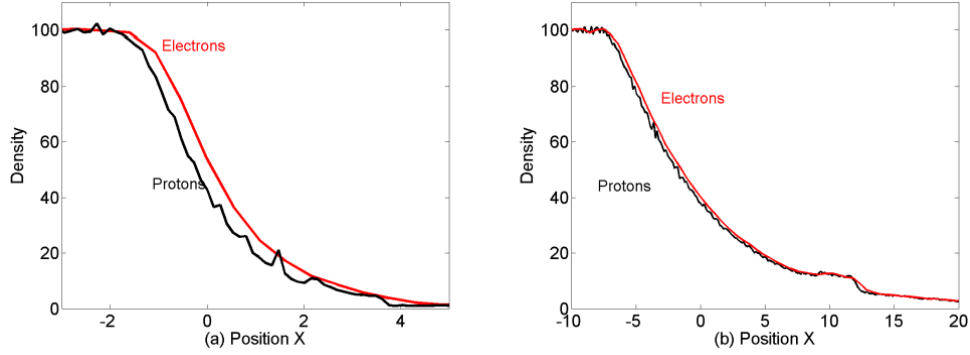}
\caption{(color online) electron and proton density of the plasma at $t_1=56$ and $t_2=288$ (frames (a) and (b) respectively); the densities are given in units of the density $n_2$ of the electrons or protons of the tenuous plasma. \label{densityplots}}
\end{center}
\end{figure*}
 The protons will react to this electric field and they will be pushed forward, resulting in the acceleration pattern visible in Fig.\ref{fig3}.b.\\
\verb|  |In the middle of this electric field structure, a sharp peak, with a width of the order of the Debye length and a relative amplitude of approximately 0.025, is formed at $x = 13$ in Fig. \ref{fig3}.c. This peak is in correspondence to a sudden decrease in the electron density (Fig. \ref{fig3}.a) and a sudden change in the proton velocity in particular of the protons of plasma 2 (Fig. \ref{fig3}.b) justifying its association with a forming shock. This is reflected also in a local increase in both the electron and proton density (Fig. \ref{densityplots}.b).\\
\verb|  |We must point out that the possibility of the proton structure in the interval $10<x<35$ and $v_x > 20 v_i$ to be a shock-reflected ion beam, as discussed in \cite{Forslund}, can be ruled out.
This is because the specular reflection of the tenuous ions should make them twice as fast as the shock and the reflected protons should keep their speed unchanged as they propagate away from the shock. We note that the rarefaction wave is a transient phenomenon, and the shock together with the shock-reflected proton beam will be established eventually.\\
\begin{figure*}[h!]
\begin{center}
\includegraphics[width=0.9\columnwidth]{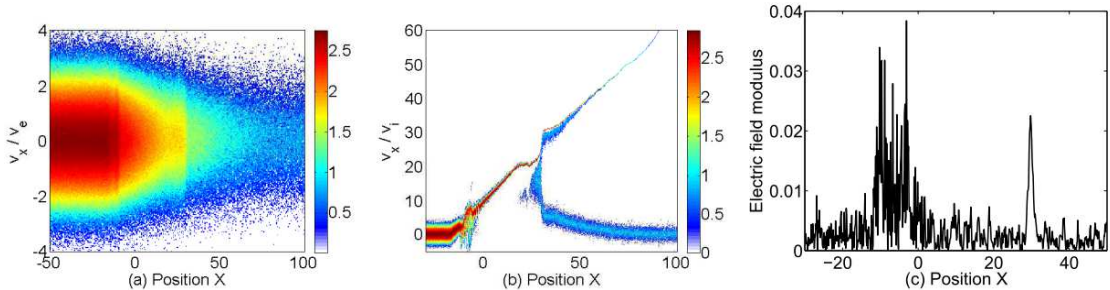}
\caption{(color online) The plasma state at the time $t_3 = 712$: Panel (a) displays the
electron distribution and (b) the proton distribution. The 10-logarithmic
color scale corresponds to the number of computational particles. The
electrostatic field modulus is shown in panel (c).\label{fig4}}
\end{center}
\end{figure*}
\verb|  |It is worth noting that, referring to typical experimental parameters (i.e. $n_{e2}\approx 10^{14}$cm$^{-3}$, $\omega_{p2}\approx5.7\cdot10^{11}$Hz), the electric field associated with the shock reads, in physical units, $E_P\approx10^7$V/m. This electric field amplitude has been demonstrated to be detectable by the probing technique and it is indeed in good agreement with recent experimental measurements (see, as an example, Fig. 1.f in \cite{Romagnani}).\\
\verb|  |Proceeding in time, the electron and proton phase space distributions, and the relative electric field distribution, at $t_3 = 712$ are shown in Fig. \ref{fig4}. The shock has now completely formed and has already propagated up to $x \approx 30$ implying a mean propagation speed of just over $0.04$. This propagation speed is comparable to the speed $20v_i \approx 0.045$ of the shock's downstream region.\\
\verb|  |This speed clearly unveils the ion-acoustic nature of the shock; the ratio between the ion-acoustic speed and the electron thermal velocity is in fact equal to $(\gamma\cdot m_e/m_i)^{1/2}$ where $\gamma$ is the adiabatic index. Generally speaking, the adiabatic index is simply a function of $f$ (the number of degrees of freedom): $\gamma = (f+2)/f$. In the case of our simulations, the protons just have one degree of freedom leading to $\gamma=3$. Given an electron to proton mass ratio of 1/1836, this ratio reads in fact \textbf{0.04}: the shock propagation speed is therefore of the order of the ion-acoustic speed.\\
\verb|  |A clarifying comment needs to be made: properly speaking, a shock demands, that the upstream flow (the plasma 2, in our case) and the plasma 1 mix in its downstream region and that the downstream plasma is heated up, as the upstream flow energy is dissipated by the shock. We see, however, that for most of the time the tenuous protons are separated from the dense ones, e.g. in Fig. \ref{fig3} at $x \approx 10$, and that the plasma heating behind the shock is negligible. Since the phase space structure resembles that expected for a shock, we henceforth will refer to it as a shock.\\
\verb|  |The rarefaction wave is still present and it looks to have expanded in the negative direction being starting now at $x\approx -16$. This implies an average expansion velocity of 0.02. This shift of the rarefaction wave towards the negative direction is apparent from the attached movie and it is associated with an increasing proton velocity at a fixed position in time.\\
\verb|  |Well behind the shock, a strongly modulated electric field, with a maximum amplitude of 0.035, is present on top of the rarefaction wave ($-10\leq x \leq0$ in Fig \ref{fig4}.c); this can be understood invoking the onset of a "drift instability" \cite{Hoh}; such an instability occurs whenever a gradient (e.g. of a density or a magnetic nature) is present within the plasma, causing a sudden disruption in the proton density that commences then to propagate in the same direction as the gradient that triggered it. This results, in the proton phase space, in a strong backward propagating wave in the dense plasma, precisely located at around $x\approx-8$ (Fig \ref{fig4}.b).\\
 \verb|  |In the meanwhile, the electrons appear to smoothly expand towards the positive direction, mainly driven by their density gradient, except from two sharp density jump located in correspondence to the shock and the wave generated by the drift instability (Fig. \ref{fig4}.a).\\
 \verb|  |At this stage of the plasma evolution, it is interesting to note that the electrostatic fields generated within the plasma appears to affect much more sensibly the protons dynamics than the electrons one. This can be explained taking into account two different reasons. First of all, the sensible difference between the initial proton and electron temperature (of 10 eV and 1 keV respectively) makes the electrostatic potential to be differently experienced by the two plasma species. Similar simulation results obtained with the same initial temperature for the two plasma species (as the one discussed in \cite{Dieckmann}) showed in fact that, in this case, also the electrons are sensibly perturbed by the electrostatic fields generated within the plasma.\\
 \verb|  |Moreover, just few electrons of the upstream region are accelerated by the electrostatic potential, because of the spatial asymmetry of the electron density distribution. The low statistical representation of a PIC simulation implies that practically no upstream electrons are seen to be accelerated. As we will discuss in detail in the following, the downstream electrons are indeed experiencing this potential whose main effect on the electron distribution is to change its shape from a Maxwellian to a flat-top one.
\section{Long Term Evolution}
\begin{figure*}[h!]
\begin{center}
\includegraphics[width=0.9\columnwidth]{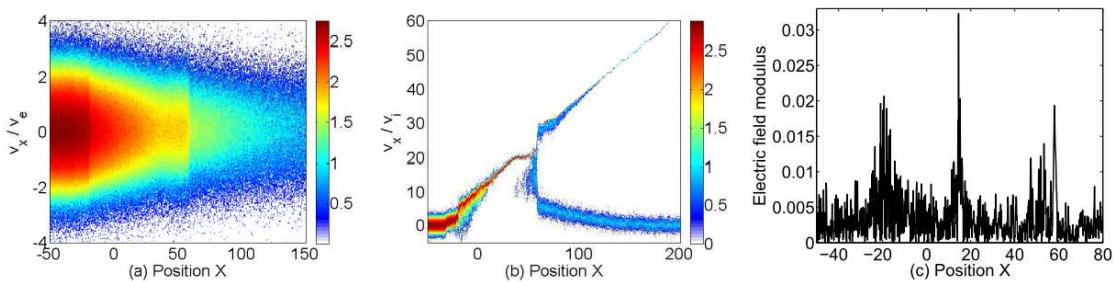}
\caption{(color online) The plasma state at the time $t_4 = 1420$: Panel (a) displays the
electron distribution and (b) the proton distribution. The 10-logarithmic
color scale corresponds to the number of computational particles. The
electrostatic field modulus is shown in panel (c).\label{fig5}}
\end{center}
\end{figure*}

\verb|  |Other two snapshots of the proton and electron phase space evolution will be then presented here in order to give an insight of the shock dynamics at later times.\\
\verb|  |In order to give a better adherence to the experimental results, we might point out here that, for typical experimental conditions in which such structures are detected (e.g. $n_{e2} \approx 10^{14}$cm$^{-3}$), the inverse of the electron plasma frequency will be of the order of $\omega_p^{-1} \approx 1.8$ps. The time steps $t_4 = 1420$ and $t_5 = 4160$, that will be discussed in the following, represent then, in physical units, 2.5 ns and 7.5 ns respectively.\\
\verb|  |The snapshots of the electron and proton phase space, together with the electric field distribution, related to $t_4 = 1420$ are shown in Fig. \ref{fig5}. The shock sustained by the rightmost electric field peak with the amplitude $\approx 0.02$ has now propagated up to $x \approx 58$ (Fig. \ref{fig5}.c), meaning an average speed of $0.04$. Behind this peak, a region of modulated electric field is present, associated with the downstream region ($45 \leq x \leq 58$) in Fig.\ref{fig5}.b.\\
\verb|  |However, the proton tip ending at $x\approx190$ has not been accelerated much beyond the $\approx 60 v_i$ it had already reached at $t=288$. The accelerating force is most likely getting weaker because it is related to the pressure gradients present at that point. Since the front of the accelerated proton beam at $x >130$ is very tenuous, no sensibly high pressure gradients can be present, implying in turn weak forces. This explains why the electric field in Fig. \ref{fig5}.c is at noise levels over wide intervals, e.g. for $20<x<40$.\\
\begin{figure*}[h!]
\begin{center}
\includegraphics[width=0.9\columnwidth]{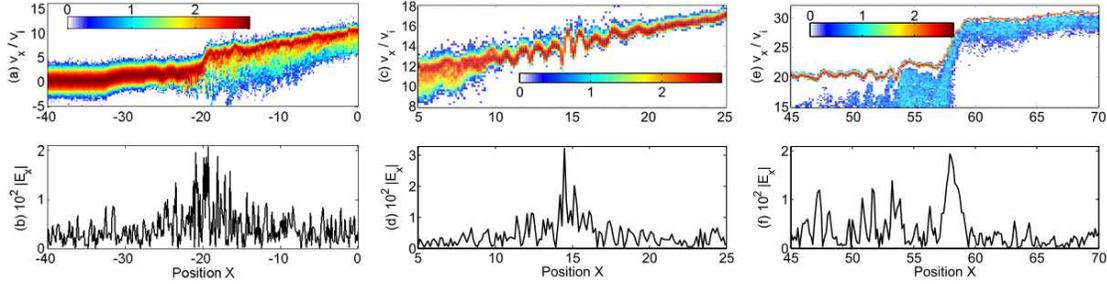}
\caption{(color online) Zoom of the 10-logarithmic proton density distribution and modulus of the electric field at $t_4=1420$ in correspondence to the start of the rarefaction wave region (a) and (b): A strong electric field, which intensity is comparable to the associated with the forward propagating shock, is associated with this wave. (c) and (d) display a rapidly oscillating wave moving along the rarefaction wave, just behind the downstream region of the shock. Finally (e) and (f) represent a zoom of the shock region; behind the peak of the electric field in correspondence to the shock, a region of modulated electric field distribution is present, related to the shock downstream region.\label{fig6}}
\end{center}
\end{figure*}
\verb|  |The rarefaction wave is still visible in the proton distribution in the intervals $-20 < x < 30$ and $100<x<190$, meaning that it kept its expansion towards the negative direction with a constant speed of 0.02. This wave is highly modulated by superposed structures. These structures, for instance the proton velocity jump located at $x\approx-25$ in Fig. \ref{fig5}.b, are driven by the drift instability as the time-animated Fig. \ref{fig2}.b demonstrates. Such a modulated region, connected with two strong electric field peaks at $x \approx -20$ and $x \approx 15$, might explain the chaotic deflection pattern, imprinted on the probing proton beam propagating through the dense plasma, seen in experiments \cite{Romagnani}.\\
\verb|  |A zoom of the relevant features outlined in this temporal snapshot of the plasma evolution are shown in Fig \ref{fig6}. They highlight the three plasma structures that give rise to strong electric fields at this time: a wave pulse at the foot of the rarefaction wave is present (panels (a) and (b)). This pulse results in proton velocity oscillations with an amplitude of about 5$v_i$ at $x=-20$ and 2$v_i$ at $x\approx -32$. The time-animated Fig. \ref{fig2}.b.) shows that it propagates to the left. A part of the rarefaction wave itself is shown (panels (c) and (d)). Here large-amplitude oscillations are observed, which have a short wavelength. These are ion acoustic waves that propagate on the rarefaction wave to the right. The proton structures in panels (a) and (c) are a consequence of the drift instability, since they can be traced back to the proton interval $-20<x<10$ that has been heated up by the drift instability. These ion acoustic waves are longlived, because the combination of hot electrons and cool ions reduces their Landau damping. This is probably the reason why they could not be observed in a simulation that used hotter ions but otherwise similar initial conditions \cite{Dieckmann}. A zoom of the main shock together with its successive downstream region is displayed in Figs. \ref{fig6}.e and \ref{fig6}.f (the reader is referred to the attached movie for a comprehensive view of the shock region). It is evident from these panels that the practically cold protons of plasma 1 are separated by a phase space gap from the protons of plasma 2. This is expected, because in a practically electrostatic one-dimensional space the particle trajectories cannot intersect in the $x-v_x$ plane. The protons of plasma 2 are heated up as they cross the shock from the right to the left at $x\approx 57$, which may be the expected shock heating. The hot and turbulent protons of plasma 2 modulate the cold dense proton beam of plasma 1 for $x<57$.\\
\verb|  |Finally, the snapshots of the plasma situation at $t_5 = 4120$ are shown in Fig. \ref{fig8}. The shock has further propagated to $x \approx 164$ without any relevant change in it shape; the propagation speed is still roughly constant around a value of 0.04.\\
\begin{figure*}[h!]
\begin{center}
\includegraphics[width=0.9\columnwidth]{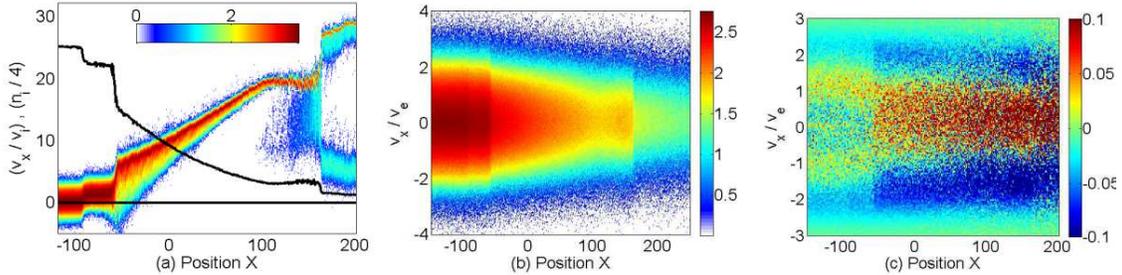}
\caption{(color online) The plasma state at the time $t_5 = 4160$: Panel (a) shows the interval of the proton distribution containing the rarefaction wave and the shock. Overplotted is the proton density in units of the ion density of
the tenuous plasma, which is divided by 4 for visualization purposes.
The 10-logarithmic color scale is given in (a) and (b) in units of
computational particles. Panel (b) show the related electron phase space; in order to better clarify the non Maxwellian behavior of it, in panel(c) the normalised distribution function $F(x,v_x)$ of the electrons is plotted on a linear scale.\label{fig8}}
\end{center}
\end{figure*}
\verb|  |In Fig. \ref{fig8}.a a profile of the proton density, given as a function of the initial proton density of the plasma 2, is overplotted. From this picture we can clearly see a structured proton density distribution with several sudden jumps that connect the high density region with the low density one.
The first two decreases, present at $x\approx-90$ and $x\approx-60$, clamp a region of uniform density and mean speed that corresponds to the strong pulse triggered by the drift instability, whose onset is already visible in Fig. \ref{fig5}.b.
After that a sudden and sensible density jump is present in correspondence to the start of the long rarefaction wave which instead is connected to a smooth and long proton density decrease. Between the rarefaction wave and the shock, i.e. at the downstream region, the proton density appears to be fairly constant.
Finally, a further sudden decrease is associated with the shock itself.\\
\verb|  |In Fig. \ref{fig8}.b the related electron phase space distribution is plotted; an interesting feature directly arising from this picture is the electron density discontinuities associated with both the rarefaction wave and the forward propagating shock. The variations in the electron density are enforced by the necessary quasi-neutrality of the plasma.\\
\verb|  |In what follows the impact of the proton expansion on the electron distribution is examined, i.e. whether or not the Maxwellian electron velocity distribution changes into a flat-top distribution, as predicted by \cite{Mora2,Grismayer}. We proceed as follows. The spatial resolution of the electron phase space data is reduced by a factor 10 to give a better statistical representation of the Maxwellian. The phase space density at each grid cell is divided by the total electron density in this cell. A constant scaling factor is multiplied to the phase space density, which ensures that the Maxwellian distribution reaches a value $1$ at $v_x=0$. This factor is computed from the initial conditions and does not take into account the changes in time of the electron phase space distribution. Finally the distribution $F(x,v_x)$ is obtained from the normalised phase space density by subtracting from it a spatially uniform Maxwellian distribution with a peak value of $1$ and a temperature of 1keV. The function $F(x,v_x)$ would show statistical flucutations at $t=0$, while trends will reveal for $t>0$ the deviation of the electron phase space distribution from the initial Maxwellian. Figure \ref{fig8}.c, displays $F(x,v_x)$ at $t_5=4160$.
\verb|  |The function $F(x,v_x)$ reveals for $-60\leq x \leq200$ a depletion of the high-energy tails, which reaches about 10\% of the peak value of the Maxwellian. The electron phase space density is enhanced at low energies. The thermal energy of the electrons has been decreased substantially by the proton expansion within the interval occupied by the rarefaction wave. For the region to the left of the rarefaction wave ($x\leq-60$), a density decrease of both the high energy tails of the Maxwellian is still visible. The much larger electron density in this interval implies, however, relatively weaker energy depletion effects.\\
\verb|  |How can we understand this electron energy loss? The electrons of the denser plasma, moving to the right, encounter the electrostatic potential at the plasma expansion front and they are reflected by it. Since the reflection occurs in the reference frame of the potential and because the potential is moving to the right, the reflected electrons will experience an energy loss in the simulation frame. Since this energy loss increases with an increasing energy of the electrons, the electron distribution is transformed to a flat-top one with a net energy loss. The practically symmetric distribution is here a finite box effect. The dense plasma covers an interval with the width $L/2=2000$, which can be crossed during the time $\Delta t=4160$ by an electron with the speed $v_e/2$. All but the coolest electrons can thus cross the interval occupied by the plasma 1 and interact with the potential at $x\approx 0$ and the one at $x\approx L/2$. The energy loss of the electrons balances the increase in the proton energy ahead of the shock. This observation is consistent with the predictions given by Mora and Grismayer \cite{Mora2,Grismayer}.\\
\verb|  |This change in shape of the downstream electron density distribution explains why the starting point of the rarefaction wave is seen to constantly propagate towards the negative direction. As firstly predicted by Mora and Grismayer \cite{Mora2}, this shift of the rarefaction wave is tightly connected with an increase of the downstream ion-acoustic speed. The ion-acoustic speed is in fact a function of the electron distribution function (see Eq. (8) in Ref. \cite{Mora2}). The progressive change in shape of the electron distribution function, that Fig. \ref{fig8}.c unveils, induces then a local change in the ion-acoustic speed that, subsequently, moves the starting point of the rarefaction wave towards the negative direction.\\
\verb|  |A summary of the acceleration that the protons in the tenuous plasma experience is shown in Fig. \ref{fig9}. The panel (a) displays the maximum proton velocity as a function of time. This picture is obtained by plotting the velocity $v_{max}>0$ corresponding to the bin that forms the upper cutoff of the proton velocity distribution. This distribution is obtained by integrating the proton phase space density from $x=-200$ to $x=500$. The panel (b) instead, shows the number of the protons with a speed $v_{max}-3v_i \le v_x \le v_{max}$ in time. From panel (a) we can clearly see that the acceleration of the fastest protons is effectively stronger at an early stage (i.e. $t<100$) whereas it decreases almost down to zero later on.
This is a further proof of the connection between the accelerating force and the density gradients: as long as a sensible gradient is present, (see Figs. \ref{fig2} and \ref{fig3}) a strong acceleration occurs whereas at later times, when the region of accelerated proton becomes more and more rarified, the acceleration flattens down to almost zero. Also the number of accelerated protons (panel (b)) further proves this hypothesis: the maximum number of accelerated protons is in fact correspondent with the inflection of the velocity function, i.e. with the point of the maximum value of the acceleration.\\
\begin{figure}[h!]
\begin{center}
\includegraphics[width=0.7\columnwidth]{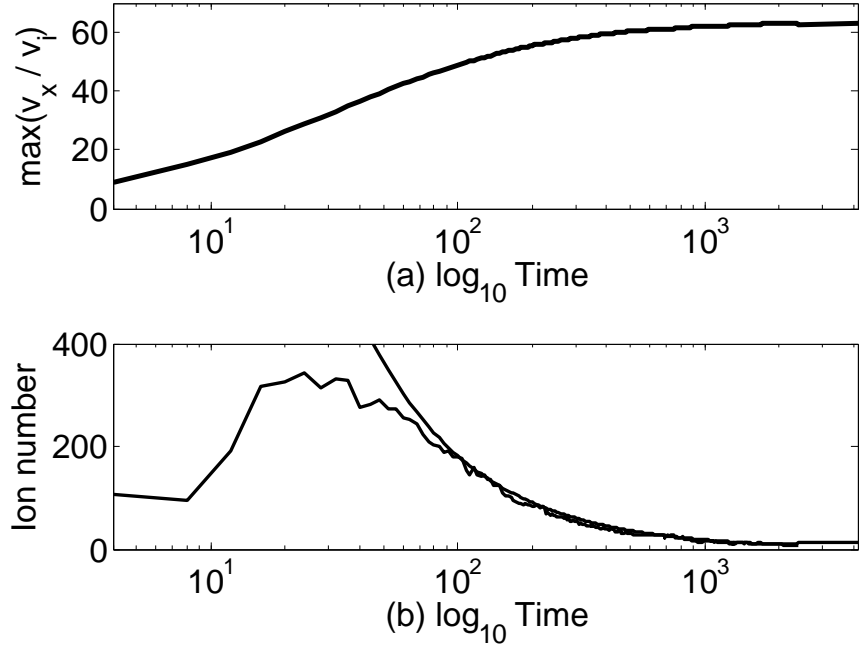}
\caption{(color online) Panel (a) displays the proton maximum velocity as a function of time whereas panel (b) shows the number of accelerated ions in time with, overplotted, a 1/t dependence for late times ($t > 10^2$).}\label{fig9}
\end{center}
\end{figure}
 After that, the number of accelerated protons decreases following a $t^{-1}$ dependence.\\
\verb|  |Such a derivation of the proton velocity and the number of accelerated protons in time has an important consequence that is worth to point out. Each bit of the proton velocity distribution, say corresponding to a velocity interval $v_0\leq v \leq v_0 + \delta v$, will be spatially stretched in time due to the different proton speeds while keeping constant the number of protons in it.
We might expect then, a density decrease induced by this stretching.
Nevertheless the integration over space exactly compensates this effect.
We can then conclude that the decrease in time of the density of the accelerated protons, following a $t^{-1}$ dependence, is entirely due to the change in the accelerating force.
\newpage
\section{Conclusions}
\verb|  |The results of 1D PIC simulations of the expansion of a dense plasma through an equally hot, more rarefied, plasma have been shown.
Both plasmas have an initial electron and proton temperature of 1 keV and 10 eV respectively; the density jump considered is sharp and with a relative amplitude of the order of 100 at $x = 0$.\\
\verb|  |The electrons of the dense plasma start immediately to diffuse, driven by their own pressure, and set a net electric field at the junction point that starts to accelerate the protons ahead of it. This electric fields eventually starts to peak up to a value of 0.025 (that, in physical units, reads $\approx 10^7$ V/m, for an electron density of about $10^{14}$cm$^{-3}$) and a width of the order of a Debye length. This electric field sits in the middle of a broader and less intense electric field distribution associated with the onset of a rarefaction wave.
Such an electric field accelerates the protons of the tenuous plasma to a maximum speed of around 60 times the initial proton thermal speed or, equivalently, 0.15 times the electron thermal speed. This acceleration is seen to saturate at later times (Fig.\ref{fig9},a), possibly due to the very low density at the tip of the accelerated protons. Over there, the electrons will maintain charge neutrality implying a very low density jump and, consequently, a very weak accelerating force. This is further confirmed by Fig. \ref{fig9}.b illustrating the number of protons accelerated by the shock as a function of time; after a very preliminary increase, the number of fast protons is seen to decrease in time following a $t^{-1}$ dependence.\\
\verb|  |Together with the shock creation, other structures are unveiled by the simulations: first of all a downstream region behind the shock and a long rarefaction wave that connects the shock with the dense plasma. The starting point of the rarefaction wave is seen to propagate towards the negative direction, possibly due to a change of the ion-acoustic speed, as predicted by recent theoretical works.\\
\verb|  |Moreover, the onset of the drift instability is clearly visible in the bulk of the dense plasma.
This instability sets the creation of secondary waves on top of the rarefaction wave that highly modulate the proton density and, consequently, the electric field distribution.\\
\verb|  |In the meanwhile the electrons progressively shift from a Maxwellian distribution to a flat-top one due to the energy loss of the high energy tail during their reflection by the moving potential barrier, in line with recently reported theoretical predictions.
This energy redistribution, leads to a net energy loss that balances the increase of energy experienced by the protons.\\
\begin{figure}[h!]
\begin{center}
\includegraphics[width=0.8\columnwidth]{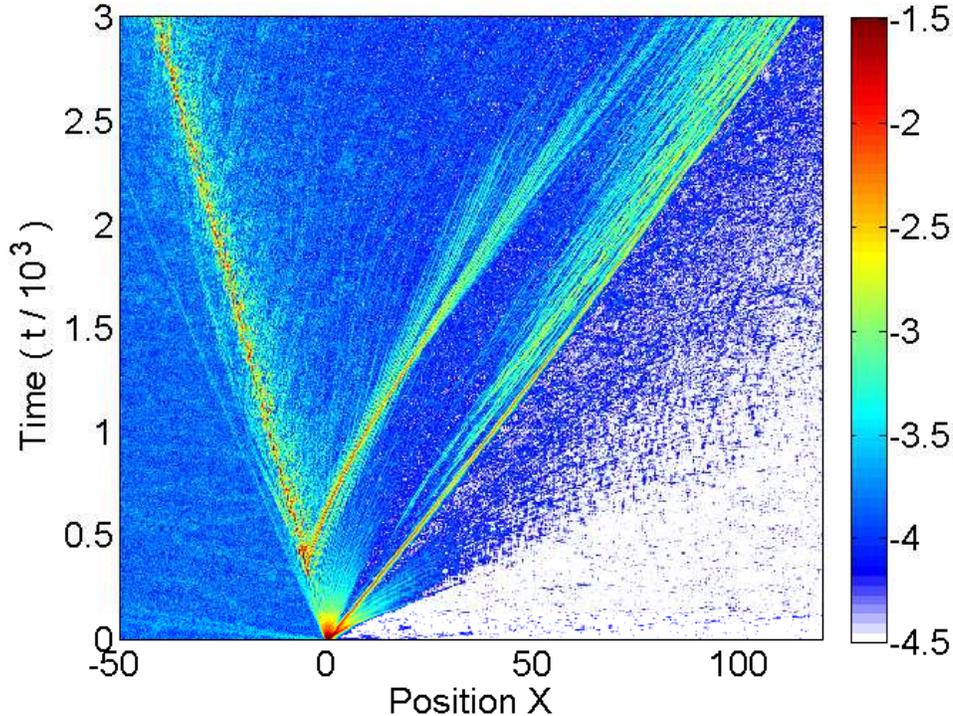}
\caption{(color online) Smoothed electrostatic field amplitude as a function of space and time. The propagation at a constant speed towards the positive direction of an electrostatic shock is clearly visible together with other secondary structures discussed in the text.}\label{efields}
\end{center}
\end{figure}
\verb|  |As a general summary of the plasma evolution and of the dynamics of the relevant structures that are triggered during such an expansion, the smoothed electric field as a function of space and time is plotted in Fig. \ref{efields}.\\
\verb|  |It is instructive to compare the results shown in this paper to the one reported in Ref. \cite{Dieckmann}, which, similarly, showed 1D PIC simulation results of the expansion of a dense plasma through a more rarefied one. The key difference is in the initial conditions set; in Ref. \cite{Dieckmann} the protons and electrons of each plasma have the same initial temperature whereas they are not thermalised throughout the entire simulation box. This different initial conditions imply a significantly different plasma evolution, the most relevant features of which are summarised below.\\
\verb|  |First of all, having initially the two plasma regions different electron densities but the same electron temperature, no double layer structure can be triggered at the interface between the two plasmas. Subsequently, no two stream instability can be generated. This yields a fundamental consequence: the electron phase space evolves smoothly in time without onset of Phase Space Electron Holes. This smooth evolution has been shown to be ideal in order to highlight the evolution of the electron distribution function towards a flat-top one.\\
\verb|  |Furthermore, the combination of hot electrons and cold ions significantly hampers the Landau damping, reducing the disturbance that this induces to ion-acoustic structures. This explains why, in our case, the shock structure is much clearer and longer lasting.\\
\verb|  |Finally, the much lower ion thermal speed allows a more efficient proton acceleration that yields to a final proton speed of the order of $\approx60$ times the initial proton thermal speed.\\
\verb|  |As an aside, we have seen that the particle acceleration witnessed in our simulations leads to a non-Maxwellian distribution function, featuring a flat-topped shape. This is in qualitative agreement (and in fact confirms) nonthermal plasma theories involving e.g., the $\kappa -$ (kappa) \cite{Hellberg} or $q-$Gaussian \cite{Tsallis,Livadiotis} distributions.\\
\verb|  |The results discussed in this paper give a further insight of the dynamics involved during the plasma expansion through a more rarefied medium, situation of relevance in both high energy density laser-matter experiments and astrophysics.
The association of these structures to sensibly high electric fields make their detection possible during astrophysical phenomena or laboratory laser-matter experiments. Further work in this direction, both from the numerical and the experimental point of view, will be focussed onto the effects that the insertion of magnetic fields will induce in the plasma dynamics and shock onset.\\
\newpage
\section*{Acknowledgments}
Funding for this research has been provided by a UK EPSRC Science and Innovation award to Queen's University Belfast Centre for Plasma Physics (Grant No. EP/D06337X/1), by GK 1203 and Vetenskapsr\aa det (Sweden).

\end{document}